\documentclass[11pt]{article}
\usepackage{blois,epsfig}

\def\Journal#1#2#3#4{{#1} {\bf #2}, #3 (#4)}

\def\APJ{{\em Astrophys.} J}
\def\MNRAS{{\em Mon. Not. R. Astron. Soc.}}
\def\NAR{{\em New Astron. Rev.}}
\def\CQG{{\em Class. Quant. Grav.}}

\def\be{\begin{equation}}
\def\ee{\end{equation}}
\def\bea{\begin{eqnarray}}
\def\eea{\end{eqnarray}}

\begin{document}
\vspace*{4cm}
\title{A CLOSER LOOK AT THE MOND NO-GO STATEMENT FOR PURELY METRIC
  FORMULATIONS}

\author{ M.E. SOUSSA }

\address{University of Florida, Department of Physics, P. O. Box 118440,\\
Gainesville, FL United States of America}

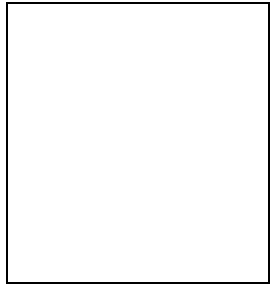
\maketitle\abstract{We reexamine the assumptions made in arriving at
  a no-go statement for purely metric formulations of MOND.  Removing
  the requirement of gravitational stability
  at appropriate scales gives life to the possibility of a purely
  metric theory of MOND.} 

\section{Introduction}
Milgrom's MOND (Modified Newtonian Dynamics) is imminently
falsifiable.  It was empirically designed to explain the fact that galactic
satellites do not experience a Keplerian fall off of their velocities
outside the central galactic bulge -- rather they asymptote to a constant
value \cite{Milgrom1}.  Therefore, it suffices to find {\em one} galaxy
which does not exhibit MOND behavior to reject it as a candidate
explanation.  

In physics, theories which can be disproved should be
welcomed.  To many MOND is a pariah, a quaint but irrelevant approach to a
phenomenon which can be described by deferring to the more ``natural'' idea of
dark matter.  This vantage point may very well be ultimately
justified, however its current acceptance is not.  Anyone who
seriously considers the problem of rotation curves must allow for
different possibilities until experimental evidence forces us to discard
any or all of them.  At this time, there has been no definitive data which can
rule out MOND.

One of MOND's greatest shortcomings has been its resistance to being
made fully relativistic.  Indeed, if we are to take it seriously as an
alteration of the gravitational force at low accelerations, it
must be somehow incorporated into a relativistic modification of
Einstein's equation.  The impetus for such a construction is partly
esthetics; however, it is phenomenology which serves as the greatest
source of motivation.  MOND in its original formulation is assumed as
an {\em alternative} to dark matter.  If so, the relativistic version
of MOND has an effect on the amount of gravitational lensing observed --
it must account for the deficiency in the General Relativity with no
dark matter prediction \cite{Mortlock}. 

As of yet, there has been no completely satisfactory relativistic
theory divised \cite{Milgrom2}.  Attempts to this end can be roughly
divided into two theoretical categories -- scalar-tensor and purely
metric.  Of course, it is often possible to write the latter
using the former.  However, the scalar-tensor models of Bekenstein,
Milgrom, and Sanders \cite{Bekenstein,Sanders} are quite
different in spirit from the purely metric one we constructed 
\cite{Soussa1}.  These particular scalar-tensor models all introduce
{\em real} degrees of freedom versus the purely gravitational degrees
of freedom of the purely metric approach.  Further, a distinction is
made between the ``Einstein'' and ``physical'' metrics.  The former is
responsible for dynamics of the gravitational field, the latter
determines the geodesics followed by test particles.  The pure metric
theory makes no distinction between the two, the {\em strong}
principle of equivalence being invoked.  

The scalar-tensor models suffered from acausal propagation of
gravitational waves, the removal of which unfortunately caused the
amount of gravitational lensing predicted to be less than that
of General Relativity alone.  One can get a
phenomenologically viable model at the expense of Lorentz
invariance by introducing a nondynamical vector field \cite{Sanders}.
The preferred-frame effects are locally suppressed; however, they are
at least as large or larger than experimental limits \cite{Sanders}.       

The purely metric theory suffers from conformal invariance of the
field equations at linearized order in perturbation theory \cite{Soussa2}.  
The result of conformal invariance in a covariant theory is the
decoupling of light.  Therefore, to linearized order, there is no
enhanced lensing, but simply the amount predicted by General Relativity.
This is the basis of our no-go statement -- and even though our
original formulation of this statement was specific to our model, it
turns out that this result is true for {\em any} purely metric theory.
The purpose here is to review the assumptions  and arguments which
lead to the no-go statement for the purely metric theory and to
comment on how one may achieve a phenomenologically viable theory of
MOND by removing the assumption of gravitational stability.

\section{The no-go statement}

Here we review the basic no-go argument previously considered
\cite{Soussa2}.  It is as follows: 
\begin{enumerate}
\item Any purely metric theory of gravity will have ten field
  equations in four spacetime dimensions of the form,
\begin{equation}
{\mathcal G}_{\mu\nu}[g]=8\pi G T_{\mu\nu},\label{gravity}
\end{equation}
where ${\mathcal G}$ (which we will call the Gravity tensor from here
on) is a function of the metric which for ordinary
General Relativity is simply the Einstein tensor.  It is obtained by
the variation of the gravitational action with respect to the metric.
$T_{\mu\nu}$ is the usual stress-energy tensor obtained from inserting
matter sources into the physical system in question.
\item The Gravity tensor is covariant by construction and thus is
  covariantly conserved.
\item The Gravity tensor can be expanded in weak-field perturbation
  theory, ${\mathcal G}_{\mu\nu}[\eta+h]$. 
\item The MOND force law scales as the {\em square root} of the mass,
\begin{equation}
F_{\rm \scriptscriptstyle{MOND}} = \frac{\sqrt{GMa_0}}{r},
\end{equation}
where $a_0$ is a constant determined by fitting to nine well measured
rotation curves -- the value of which is $(1.20\pm .27)\times
10^{-10}$ m s$^{-2}$.  This follows from the nonrelativistic MOND
force law \cite{Milgrom1} which gives the required constant asymptotic
velocities of satellites in circular (or nearly circular) orbits.
\item In the deep MOND regime (that is, for accelerations on the order
  of $a_0$), at least one component of $h_{\mu\nu}$
  must scale as $\sqrt{GM}$ if rotation curves are to be reproduced.
\item The right hand side of (\ref{gravity}) scales as $GM$. 
\item From 6. it must be that in the deep MOND regime at least one of
  the ten equations is non-zero at order $h^2$.
\item A second rank symmetric tensor in four dimensions has two
  distinguished components -- its covariant derivative and its trace.
  The first is zero by 2.  Therefore, the component which begins at
  quadratic order in the weak-fields in the deep MOND regime is the
  trace component. 
\item The {\em linearized} MOND weak-fields are thus traceless.
\item Tracelessness directly implies conformal invariance and
  henceforth the decoupling from light.
\end{enumerate}
This last statement essentially kills any hope of formulating a
phenomenologically viable theory, for the linearized MOND equations
give no added lensing to the result of General Relativity given the
assumptions of the next section.  

\section{The Assumptions}

The assumptions made in the previous section were \cite{Soussa2}: 
\begin{enumerate}
\item The gravitational force is carried by the metric with its source
  being the usual stress-energy tensor.
\item Gravity is described by a covariant theory.
\item The MOND force law can be realized in weak-field perturbation
  theory.
\item The theory of gravity is absolutely stable
\item Electromagnetism couples conformally to gravity.
\end{enumerate}
The third and fifth assumptions are the most rigid.  The third, if not true,
would bar us from working with any relativistic version of MOND -- if
there is no region for which the MOND force is weak (or at least as 
weak as the Newtonian gravitational force), then there is no hope in 
passing standard phenomenological requirements.  

The first, second, and fourth assumptions, however, are capable of
undergoing more scrutiny.  The first, for example, may be violated if
one makes a distinction between a ``physical'' and ``gravitational''
metric.  In such a case particles would follow geodesics of the former
while gravity would behave according to dynamics of the latter.  

The second assumption is easily foregone if one specifies a
preferred-frame.  This can be accomplished by inserting
nondynamical vector fields \`{a} la Sanders \cite{Sanders}.  As
mentioned earlier, the effects coming from preferred-frame physics
would need to be suppressed at least enough to pass local
gravitational tests.  

These already mentioned assumption violations, however, do not bear
very heavily on a purely metric formulation.  The first applies if one
wishes rather to consider scalar-tensor theories.  The second is
contrary to the philosophy of a pure metric approach -- that is, we do
not want to introduce any new degrees of freedom, regardless of
whether they are non-dynamical.  The fourth, however, is the most
significant to examine if we are interested in remaining close
to the original idea of pure metric theories.   

Given the choice between a stable and unstable theory, the physicist
will always choose the former.  However, when doing phenomenology, the
latter may be the choice of greater utility.  When the instability
manifests itself at scales outside or nearly outside the physical
scale, the phenomenologist may cautiously accept (or at least consider
accepting) the unstable solution as viable.  

It is possible to imagine that {\em all} of
the linearized MOND weak-fields vanish in the equations of motion, in
which case there would no longer be a linearized theory,  and
sufficient bending
of light could be realized.  If the MOND weak-fields
begin at quadratic order in the field equations, they must be cubic in
the action and thus possess an inherent instability.  This is not
necessarily a fatal property.  There are {\em two} weak-field regimes
-- the deep MOND (or ultra-weak-field) and the weak-field (or
Newtonian).  In regions such as the solar system it would be the
Newtonian regime which dominates and thus we would experience no
deviation from well established physics.  At larger scales (galactic
and/or cosmological) we would expect the deep MOND regime to enter the
fold.  The unstable solution would decay then into large wavelength
particles (galactic scale at least) diffusing as the universe
expands.  This process could directly result in the return to the
Newtonian regime as decay products would build a sufficiently large
gravitational potential.

\section{Conclusion}
To date there has not been a completely satisfactory theory of MOND
which is both fully relativistic and phenomenologically viable.  The
pure metric approach will fail to pass the crucial test of
gravitational lensing due to the conformal invariance of the
linearized weak-field equations in the deep MOND regime.  At the cost of
stability, one restores hope that the theory predicts
the correct amount of observed lensing.   This instability is
phenomenologically acceptable if its behavior is detectable at least
at galactic scales.  We may conclude that although a purely metric
approach has some serious issues with which to contend, it is not to
be considered dead and gone.  If one divises a theory in which all ten
of the linearized field equations vanish then there is hope for
sufficient lensing.  Further, a no-go statement as the one we
have made should be taken as a challenge.  It
only takes one counter example to disprove the statement, just as it
takes only one rotation curve which contradicts MOND to reject it as a
candidate explanation to one of the most interesting problems
currently facing the astrophysics community.     

\section*{Acknowledgments}
I would like to thank R.P. Woodard for his encouragement and support
during the completion of this project.

\end{document}